# A Study on Quantum Neural Networks in Healthcare 5.0

Sanjay Chakraborty

*Abstract*—The working environment in healthcare analytics is transforming with the emergence of healthcare 5.0 and the advancements in quantum neural networks. In addition to analyzing a comprehensive set of case studies, we also review relevant literature from the fields of quantum computing applications and smart healthcare analytics, focusing on the implications of quantum deep neural networks. This study aims to shed light on the existing research gaps regarding the implications of quantum neural networks in healthcare analytics. We argue that the healthcare industry is currently transitioning from automation towards genuine collaboration with quantum networks, which presents new avenues for research and exploration. Specifically, this study focuses on evaluating the performance of Healthcare 5.0, which involves the integration of diverse quantum machine learning and quantum neural network systems. This study also explores a range of potential challenges and future directions for Healthcare 5.0, particularly focusing on the integration of quantum neural networks.

## I. INTRODUCTION

**A**n artificial neural network (ANN), known as a quantum neural network (QNN) uses the concepts of quantum computing to perform computation. It combines the principles of neural networks and quantum computing to offer improved capabilities for some applications. A quantum neural network employs qubits as the fundamental unit of information in contrast to a classical neural network, which processes and represents information using classical bits. Qubits can represent multiple states simultaneously due to their ability to exist in superposition. This characteristic enables quantum neural networks to concurrently process and analyze data across different quantum states, potentially leading to exponential acceleration of computation in certain scenarios. Quantum gates are utilized to manipulate qubits and perform computations, while measurement operations are employed to extract information from quantum states, forming the fundamental architecture of a quantum neural network [1]. Both the quantum variational algorithm and the classical gradient descent training algorithm can be employed to train quantum neural networks. Quantum neural networks hold significant potential for various industries, offering applications in tasks such as pattern recognition, classification, and regression in the realm of quantum machine learning [29]. Furthermore, quantum neural networks can be utilized for quantum simulations, quantum chemistry, and optimization. However, it is important to note that practical implementations of quantum neural networks face significant challenges due to existing limitations of quantum hardware, including noise, decoherence, and constrained qubit connectivity. In order to build scalable and fault-tolerant quantum computing systems that can effectively leverage the capabilities of quantum neural networks, researchers are actively exploring strategies to tackle these challenges. Quantum neural networks, situated at the intersection of quantum computing and artificial intelligence, are an intriguing area of study that holds the potential to unlock new pathways for addressing complex computational problems [8,83]. Artificial neural networks play a crucial role in deep learning algorithms. Fig 1 shows different applications of quantum computing in the healthcare domain [46,52]. Due to the advancements in

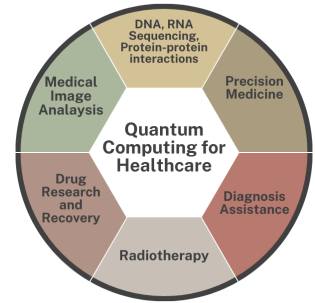

Fig. 1: Quantum computing applications in Healthcare

classical deep learning and the progress made in quantum computing, quantum neural networks (QNNs) have gained increasing popularity. QNNs are similar to conventional neural networks but incorporate variational parameters, which allow for greater flexibility in modeling and optimization [2]. There are several reasons to build a quantum version of neural networks. First, there are a number of ways in which quantum computers can outperform conventional computers. Certain algorithms based on the quantum Fourier transform, such as Shor's algorithm, have demonstrated the potential for achieving quadratic speedups compared to the most popular classical approaches [57]. Additionally, it has been shown that specific quantum resources, including quantum nonlocality and contextuality, can provide unconditional advantages in solving certain computational problems. These findings highlight the

Sanjay Chakraborty is working in the Department of Computer and Information Science (IDA) at Linköping University, Sweden and Department of Computer Science & Engineering, Techno International New Town, Kolkata, India (Email: schakraborty770@gmail.com).

unique capabilities of quantum computing in tackling complex computational tasks. These intriguing findings provide impetus for further research into the potential benefits of quantum neural network (QNN) models, particularly in the context of big data. A second rationale for utilizing QNN models emerges when attempting to extract knowledge from a quantum dataset, as opposed to a conventional dataset. In such scenarios, employing a quantum model to tackle the task becomes more sensible [3]. Extracting sufficient details from a quantum state for a classical device becomes increasingly challenging as the system scales up. However, a QNN model, with its ability to naturally handle data in the exponentially large Hilbert space, could prove advantageous [36]. Lastly, preliminary studies suggest that quantum neural networks, compared to similar feedforward networks, have the potential to achieve higher useful dimensions through faster training. The effective dimension is a measure of a model's ability to generalize well on new data [4]. In the context of Healthcare 5.0, quantum deep neural networks (QDNNs) have the potential to revolutionize the analysis of medical data and decision-making [59]. QDNNs can effectively handle vast amounts of medical data, including genomics, patient records, and medical imaging, by harnessing the computational power of quantum computing [53]. This capability enables the network to identify intricate patterns and correlations that may go unnoticed by traditional methods [14]. As a result, patient outcomes are enhanced through more accurate disease diagnosis, personalized therapy recommendations, and medication discovery. Additionally, QDNNs offer enhanced privacy protection through the use of quantum cryptography, ensuring secure sharing of sensitive medical data [60]. However, for QDNNs to fully unleash their potential in advancing precision medicine and healthcare delivery, they must overcome technological challenges and adhere to regulatory requirements before practical integration into healthcare systems can be achieved [15, 28, 44]. There are several other variations of the QNN, such as the random QNN, which is used for recognizing noisy images [35], the recurrent QNN [37, 77], and quantum deep reinforcement learning [70, 73], among others. In this study, we argue for further exploration of quantum computing in Healthcare 5.0. We begin by providing background information on quantum neural networks and healthcare 5.0. Then, we challenge common misconceptions about them in relation to our main research themes. To gain a better understanding of how practice studies contribute to our knowledge of QNN in the Healthcare Industry 5.0, we have developed a roadmap that incorporates important themes and identified research gaps into a matrix of research questions. Practice studies have the potential to offer insights into the real-world impact of QNN-supported healthcare analytics, going beyond traditional quantum computing research. We also highlight relevant research gaps that can be addressed through practice investigations. Table I provides a yearwise classification of quantum computing applications to healthcare 5.0. The study's conceptual framework is shown in Fig. 2.

*A. Paper Organization*

In the following sections of this study, we will delve into the background and principles of quantum computing, explore variational quantum solvers, and elucidate the typical architecture of quantum neural networks. We will then examine various quantum neural network systems and their applications in different domains of healthcare 5.0. Additionally, we will provide a comprehensive comparison, discuss various challenges and explore further research avenues.

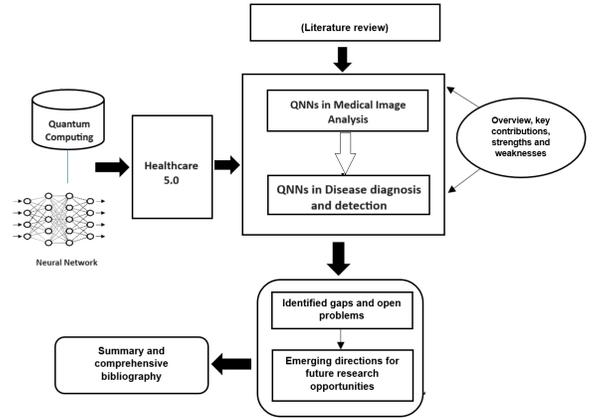

Fig. 2: Conceptual framework of the study

## II. BACKGROUND

*A. Quantum Computing*

Compared to classical computing, quantum computing operates and manipulates information in fundamentally different ways by leveraging the principles of quantum physics. Classical computers rely on bits, which can represent information as either 0 or 1. However, quantum computers employ quantum bits, or qubits. One of the key features of qubits is superposition, which allows them to exist in multiple states simultaneously. This unique characteristic enables quantum computers to perform calculations in parallel, leading to significantly faster solutions for certain problems compared to conventional computers. Qubits can also exhibit entanglement, which means that the states of two qubits can be interdependent even when they are physically separated by long distances. This unique property allows quantum computers to perform sophisticated computations and tackle problems that are beyond the capabilities of classical computers, such as simulating quantum systems or factoring large numbers [7]. However, there are several challenges that need to be addressed before functional quantum computers can be developed and operated. These challenges include ensuring qubit stability, mitigating decoherence, and implementing error correction techniques [18]. However, ongoing research and advancements in quantum computing have the potential to revolutionize various industries, such as medicine development, materials science, cryptography, and optimization. Quantum computation, unlike classical bits, relies on the concept of qubits [30]. Through the unique properties of qubits, quantum computers can offer unprecedented computational power and capabilities, leading to breakthroughs in solving complex problems and driving innovation across multiple sectors. In binary informa-tion processing, superpositions of $|0\rangle$ and $|1\rangle$ qubit states are employed in addition to information storage, such as,

$$|\psi\rangle = \alpha|0\rangle + \beta|1\rangle \quad (1)$$

where $|\alpha$ and $\beta$ denote probability amplitudes of the basis state in a Hilbert space ($H = C^2$), and if $|\alpha|^2$ is larger than $|\beta|^2$, then the probability of preserving the information is greater to the $|0\rangle$ basis state, and vice versa. Thus, we may express the superposition state $|\psi\rangle$ as follows,

$$|\psi\rangle = \sum_{i=0}^{2^n-1} \alpha_i |i\rangle \quad (2)$$

where the chance of receiving value $|i\rangle$ is represented by $|\alpha_i|^2$. All the binary integers from the set $0, 1, ..., i, ..., 2^n - 1$ with the accompanying probability $|c_i|^2$ represent the n qubit quantum register in equation (2). Another geometric representation of a qubit is a three-dimensional Bloch sphere. However, we can also rewrite equation (1) as,

$$|\psi\rangle = e^{i\gamma}(cos\frac{\theta}{2}|0\rangle + e^{i\phi}\sin\frac{\theta}{2}|1\rangle) \quad (3)$$

where, $\theta, \phi$ and $\gamma$ are real numbers. One possible rewrite of the same equation is,

$$|\psi\rangle = cos\frac{\theta}{2}|0\rangle + e^{i\phi}\sin\frac{\theta}{2}|1\rangle \quad (4)$$

TABLE I: Year-wise (2023-2017) classification of works on Quantum Computing in Healthcare 5.0

| Key Research Areas | Year | Applications |
| --- | --- | --- |
| Deep quantum computing circuit embedded in a machine learning architecture [51] | 2023 | predict the drug response for cancer patients |
| Quantum Deep Potential Interaction Model [54] | 2023 | Drug discovery |
| Quantum learning inspired by AlexNet [63] | 2023 | Categorization of Brain disorder symptoms |
| Quantum fruit fly algorithm inspired by ResNet50-VGG16 [64] | 2023 | Categorization of medical diseases |
| Transfer-learning-based deep learning protocol [65,71] | 2023 | image classification, Protein calculations for high precision prediction in large chemical and biological systems |
| Quantum support vector classifiers (QSVC) and variational quantum classifiers (VQC) [68,88,96] | 2024 | Breast cancer detection [68], Heart Failure detection [88], smart healthcare [96] |
| Quantum orthogonal neural networks [26,75] | 2021 | Medical image analysis |
| MRI-radiomics variational Quantum Neural Network [76] & quantum annealing [89] | 2023 | medical image analysis [76], Differentiate Between Large Brain Metastases and High-Grade Glioma [89], Respiratory disease detection [95] |
| Quantum Convolutional Neural Networks [78,91,97] | 2024 | detecting pneumonia from chest radiographs [78], Brain Tumor diagnosis and biomedical image classification [91,97] |
| Quantum precision-based granular approach [98], Quantum-Inspired Self-Supervised Network [102] | 2020, 2023 | Segmentation of brain MRI |
| Hybrid classical–quantum transfer learning [47,81] | 2022 | detection of Alzheimer's |
| Quantum support vector classifiers [82], IoMT based quantum blockchain and quantum neural network [93] | 2023 | Abnormal heart-beat detection |
| Quantum Single Layer Perceptron [84] | 2022 | Classification of real-world data |
| Quantum Generative Adversarial Networks [25,85] | 2023 | Medical image augmentation |
| Heap Based Optimization with Deep Quantum Neural Network [86] | 2023 | identifying and classifying healthcare data |
| Quantum photonic convolutional neural network [87, 90] | 2023 | Secure AI-inspired healthcare systems |
| Quantum-LSTM contrastive learning [92] | 2023 | Mental health monitoring |
| Blockchain-inspired quantum heuristic algorithm [94], Quantum Relu activation function for Convolutional Networks [58,101], 3D Quantum-inspired Self-supervised Tensor Network [103] | 2022, 2023 | secure healthcare prediction [94], Parkinson's disease diagnosis[101], Volumetric segmentation [103] |
| Quantum optical neural network classifier [104] | 2024 | Smart healthcare application |
| IoTs-spiro system and fuzzy-based quantum neural network classifier [105] | 2023 | Chronic obstructive pulmonary disease prediction |

On the 3-D Bloch sphere, a point is defined by the numbers $\theta$ and $\phi$. The pure states of the system are represented by the points on the Bloch sphere's surface, while the mixed states are represented by the points inside. The probability for quantum state $|\psi\rangle$ is being represented using Dirac notation and computed as $|\langle\phi|\psi\rangle|^2$. The remaining quantum operations are expressed by different quantum gates. The quantum gates are all reversible and unitary in nature ($U.U^\dagger = I$, U=unitary operator, $U^\dagger$=inverse of U, I=identity matrix).

### B. Quantum Neural Networks

To enhance processing capacity and address complex tasks more efficiently, quantum machine learning [79] integrates quantum computing principles into the field of machine learning [74]. This approach involves the use of quantum neural networks (QNNs), which leverage quantum computing concepts like superposition and entanglement [6, 38] to process input data and perform computations. Unlike traditional neural networks that utilize classical bits [48, 49], QNNs employ qubits for information processing. This enables QNNs to effectively handle large datasets and potentially achieve enhanced performance in tasks such as classification, regression, and optimization [19]. The field of quantum machine learning holds promise for applications in various domains, including financial modeling, drug discovery, and optimization [11, 13]. However, extensive research and advancements are still required to fully harness this potential and overcome existing technical challenges. The fusion of entanglement, inference, and natural parallelism in quantum computation has led to the development of artificial neural networks with quantum characteristics [16, 20]. In the realm of quantum computing, a variational quantum circuit (VQC) is a flexible and promising technique that applies the principles of variational approaches to optimization problems. It seeks to identify optimal solutions for specific computational challenges by adapting a parameterized quantum circuit to minimize a designated cost function [43, 61]. The adaptability to diverse problem domains and compatibility with existing classical optimization techniques are among the notable advantages of VQCs.

These circuits are instrumental in quantum machine learning [62], quantum chemistry simulations [66], and optimization problems, showcasing their potential to drive technological advancements across various domains. By harnessing the power of quantum computation, VQCs offer a precise and efficient means to tackle complex challenges. Additionally, the Variational Quantum Eigen Solvers (VQE) [17] approach, which integrates quantum and classical machine learning, represents a significant development in the field. VQE combines the strengths of quantum and classical computing to solve complex eigenvalue problems, further expanding the potential applications of VQCs. The objective of the Variational Quantum Eigen Solvers (VQES) approach is to determine the lowest eigenvalue of a quantum Hamiltonian matrix H. VQES-QNNs efficiently approximate the ground state energies and associated eigenstates of quantum systems by combining the computational power of variational algorithms with the flexibility of neural network architectures. These hybrid methods leverage the trainable parameters within the quantum neural network to dynamically optimize quantum circuits, facilitating the resolution of various quantum mechanical problems and exploration of complex Hilbert spaces. This approach opens up possibilities for state-of-the-art computational modeling in fields such as drug development, molecular simulations, and personalized medicine, which have the potential to revolutionize the healthcare industry. VQES-QNNs have the capability to accurately predict molecular properties, optimize drug molecules, and analyze intricate biological systems with remarkable efficiency [50]. This accelerated computational approach plays a crucial role in expediting the drug development process [51] and facilitating the discovery of novel therapeutics. Additionally, by simulating biological processes at the quantum level, these hybrid techniques have the potential to enhance treatment plans and diagnostics, leading to more personalized and effective medical interventions. A variational quantum circuit is represented in Fig 3.

Variational Principle: $|\lambda_{min}\rangle \leq \langle H \rangle_\psi$

Fundamental Steps:
1. Map the molecular Hamiltonian into a qubit Hamiltonian.
2. Prepare the quantum ansatz $|\psi(vec(\theta))\rangle$.
3. Measure the expectation value of $\langle \psi(vec(\theta)H|\psi(vec(\theta))\rangle$ using a classical optimizer (with a varying ansatz).
4. Iterate the process until convergence.

A quantum neural network is a quantum counterpart of a traditional network design, where the principles of quantum computing and deep learning are combined to process and encode information. Quantum deep neural networks (QDNNs) leverage the superposition and entanglement capabilities of qubits to perform computations and store data [9, 41]. There are three approaches to integrating data and computation: classical data with quantum computation, quantum data with classical computation, and quantum data with quantum computation. Quantum dot net networks (QDNNs) utilize quantum circuits for computational tasks and find applications in quantum chemistry, quantum machine learning [80], and optimization. While quantum neural networks have the potential to solve complex problems more efficiently, their scalability and practical implementation pose significant challenges [22, 23]. To overcome the current limitations of hardware, these networks often operate in a hybrid fashion, combining both quantum and classical computing approaches [39, 40, 42]. This hybrid quantum-classical methodology allows for the utilization of existing classical computing resources while leveraging the unique capabilities of quantum systems. By integrating the strengths of both quantum and classical computing, researchers can work towards overcoming the hardware restrictions and advancing the practical applications of quantum neural networks. A typical workflow of a quantum deep neural network is given below,

**Step-1**: A neural network model can be built with trainable parameters ($\theta$).

$$f_{NN}(x;\theta) = \sigma(wx+b), \theta = w, b \quad (5)$$

**Step-2**: Derive a cost function ($C$),

$$C = \sum_{data x,y} |f_{NN}(x;\theta) - y|^2 \quad (6)$$

**Step-3**:
a. Compute the gradient of cost (C) concerning parameters ($\theta^{(t)}$) [12].
b. Update parameters ($\theta^{(t+1)}$) in the direction of the gradient ($\nabla$).
c. Repeat until convergence.

$$\theta^{(t+1)} = \theta^{(t)} - \eta \nabla_\theta C \quad (7)$$

where, $\eta$ is the learning rate. **Step-4**: After the training phase, prediction and classification can be performed using the model.

The basic building steps of another widely used quantum network called the quantum convolutional neural network (QCNN) [10,55,56], which is suitable for medical image analysis are given below,

**Step-1**: The encoding of given classical data is needed where classical data are transformed into single or multi-qubit states.

$$Class_{Data} \rightarrow |\psi\rangle = \alpha|0\rangle + \beta|1\rangle \quad (8)$$

**Step-2**: The transformed quantum state $|\psi\rangle$ is subjected to a sequence of quantum gates [32]. The quantum convolution is the result of this process.

**Step-3**: A controlled swap gate (CSWAP) now handles the quantum pooling process [5]. The quantum Hadamard gate is denoted by H.

$$H|0> \rightarrow CSWAP(|0>, |\psi\rangle) \rightarrow H \quad (9)$$

**Step-4**: This layer is composed of feed-forward oriented neurons, meaning that all the upcoming neurons are coupled to all the antecedent neurons [31].

**Step-5**: Apply the measurement (M) of the resultant quantum states ($\phi$) through a classical binary register.

$$M = <\phi''|Z|\phi''> \quad (10)$$

The first qubit is used to measure the expected value of the Pauli-Z operator during this phase.

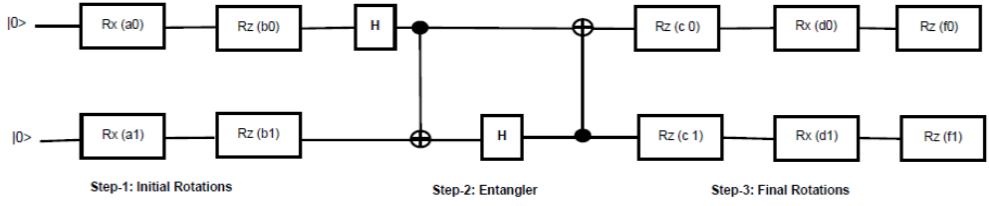

Fig. 3: Quantum variational eigensolver circuit

## C. Healthcare 5.0

In order to build a highly customized, effective, and patient-centered healthcare system, Healthcare 5.0, sometimes referred to as Smart Healthcare, combines cutting-edge technology including artificial intelligence (AI), the Internet of Things (IoT), robotics, big data, and blockchain. It enhances patient outcomes, treatment plans, and diagnostics by utilizing automation, real-time data, and advanced analytics. In order to improve healthcare quality and lower costs, the goal is to shift from traditional reactive care to a proactive and predictive model where AI and connected devices help medical personnel provide preventive care, remote monitoring, and personalized treatment [86], [87].

## III. QUANTUM NEURAL NETWORK IN HEALTHCARE 5.0

The capability of quantum neural networks (QNNs) to tackle complex data and perform tasks such as drug discovery, medical image analysis, and patient diagnosis at a faster pace compared to traditional approaches has generated significant interest in the healthcare industry. While the healthcare domain has traditionally relied on classical approaches, the contribution of the quantum realm is also making notable strides in this field [99]. Table II provides a taxonomic classification of a set of key quantum techniques in Healthcare 5.0.

### A. QNNs in Medical Image Analysis

One study introduced a unique three-dimensional quantum-inspired self-supervised tensor neural network (3D-QNet) for volumetric segmentation of medical images with the advantage of requiring less training and supervision. This topology enables the processing of 3D medical image data voxelwise, making it especially appropriate for tasks involving semantic segmentation. Each volumetric layer makes use of quantum neurons, also known as qubits or quantum bits [34]. Quantum rotation gates are used to describe the interconnection weights and the input quantum neurons store the intensity of pixels as qubits. Compared to conventional supervised and self-supervised networks, the network demonstrates faster convergence due to the integration of tensor decomposition within the quantum formalism. A mapping method converts the classical intensity of each $i^{th}$ normalized grayscale image pixel of an MR or CT volume ($\alpha_i$) into a quantum state. This research incorporated several innovative elements, including the utilization of nonlinear transformations, neighborhood-based topological interconnections among network layers, and parametrized Hadamard gates. These additions, along with the foundation of the qutrit framework, formed the basis of the proposed model. The evaluation of the model on Berkeley gray-scale images demonstrated its satisfactory performance, showcasing the effectiveness of the proposed QFS-Net [103].

$$Fn(\alpha_i) = [cos(\frac{\pi}{2}\alpha_i) sin(\frac{\pi}{2}\alpha_i)] \quad (11)$$

The quantum interconnection weight can be represented as,

$$|Wt(\omega_{i,j}) >= [cos(\frac{\pi}{2}\omega_{i,j}) sin(\frac{\pi}{2}\omega_{i,j})] \quad (12)$$

In a study conducted by researchers [89], an MRI-radiomics variational quantum neural network (QNN) was employed to construct a brain tumor model using an MRI dataset. To address this task, the researchers utilized the mutual information feature selection (MIFS) method, which transformed the problem into a stochastic optimization task. This task was subsequently solved using a quantum annealer. Notably, the method exhibited strong performance and achieved accuracies comparable to those of conventional approaches. A Quantum-based self-supervised network model designed for brain MRI segmentation is presented in this paper. Using the ideas of quantum computing, this novel method creates a self-supervised architecture intending to simplify the segmentation procedure without requiring considerable manual supervision. The input, hidden, and output layers of the QIS-Net architecture are characterized by qubits and are made up of three layers of quantum neurons. In the $t^{th}$ sample sets at the network layer of the proposed model, the initial input–output relation of a $j^{th}$ basic quantum neuron can be expressed below,

$$|X_j^t\rangle = \sigma_{QISNET}(\sum_k^{n \times m} f(|X_j^{t-1}\rangle\langle\phi_k^t|\zeta_k^t\rangle)) \quad (13)$$

where $|X_j^t\rangle$ denotes the output of the quantum neuron j, and $|\zeta_k^t\rangle$ is the threshold parameter. The activation function known as quantum-inspired multi-level sigmoidal activation ($\sigma_{QISNET}$) is described as follows,

$$\sigma_{QISNET}(g) = \frac{1}{\lambda_\omega + e^{-v(g-\eta)}} \quad (14)$$

where v=steepness factor and $\eta$ is an activation value represented by qubits and $\lambda_\omega$ represents the multi-level class response. Compared to conventional supervised methods, the model shows improved convergence and efficiency by utilizing quantum-inspired methodologies. Promising results from extensive testing and evaluation of brain MR image datasets highlight the promise of this quantum-inspired method for precise and effective brain image segmentation tasks [102]. Notable efficiency has been demonstrated by quantum neural networks in the diagnosis of pneumonia and brain tumors [78,91,97]. This study used an open-source brain tumor image collection to classify brain tumors using QCNNs. Although not groundbreaking, this work presents a number of state-of-the-art developments and achieves better results in both out-of-distribution generalizability and in-distribution categorization. The basic steps of QCNNs are described in section II [91,97]. Using six thousand four hundred labeled MRI scans with 2 classes, a hybrid model is provided for the diagnosis of Alzheimer's. The use of hybrid classical–quantum transfer learning allows for the best possible pre-processing of large-dimensional, complicated data. Extracted large-dimensional features by ANN are subsequently integrated into a quantum processor as significant vectors of features. Resnet34 is applied to fetch features from the given image, and then load a five-hundred-twelve-feature vector into the developed QVC to obtain a 4-feature vector for exact decision boundaries. Furthermore, some renowned quantum simulators are used to distinguish between demented and non-demented images to validate the model. Retaining all of the convolutional and pooling layers in place, ResNet34 is utilized as an already trained model, and a fresh trainable classifier is initially incorporated into the network to categorize the MRI images. This method is called classical-to-classical transfer learning. In the following stage, classical-to-quantum transfer learning is applied. ResNet34 extracts the features and a trained quantum circuit is applied for classifying MRI images [81]. A study [63] suggested using magnetic resonance imaging (MRI) data to identify neurological conditions using an AlexNet-inspired quantum transfer learning technique. In this research work, a new hybrid model is created by combining CNN's classical layers with variational circuits' quantum layers (QVC) and then adding a fully connected layer on top. Convolutional layers, nonlinear Relu activation function, and adaptive pooling layers—which serve as feature extractors and offer significant patterns—are the building blocks of classical layers. A few score vectors are intended to be created from the rich information found in a large amount of data by the feature extraction algorithms. This work uses the CNN architecture known as AlexNet. This model serves as an extractor of features, and QVC builds a quantum circuit using qubits and quantum gates to gain an acceleration in processing demands. To reduce the overfitting and imbalanced data problems, a research work was introduced. This work employs the 'Quantum Fruit Fly Algorithm (QFFA)' technique to enhance the efficacy of classification in medical diagnosis through feature selection. The images are normalized using the Min-Max Normalisation approach. For feature extraction, the deep learning models ResNet50 and VGG16 were used. The SVM model was fitted with characteristics selected using the QFFA technique to effectively classify medical conditions [64].

TABLE II: Taxonomic classification of key quantum techniques in Healthcare 5.0

| Main Research Areas | Key techniques for healthcare |
|---|---|
| Quantum Computing | Quantum precision-based granular approach [98], Quantum annealing [89], Quantum blockchain [93] |
| Quantum machine learning | graph convolutional, convolutional, and quantum layers embedded in a machine learning architecture [51], Quantum Fruit Fly Algorithm [64], Quantum support vector classifiers (QSVC) [21] and variational quantum classifiers (VQC) [96], quantum heuristic algorithm [94] |
| Quantum Deep Neural Networks | Quantum Deep Potential Interaction Model [45,54], Quantum Learning through Alex-Net [63], Quantum convolutional neural networks [78,91,97], Quantum orthogonal neural networks [75], MRI-radiomics variational Quantum Neural Network [76], Quantum-LSTM contrastive learning [92], Hybrid classical–quantum transfer learning [72,81], Quantum Single Layer Perceptron [84], Quantum photonic convolutional neural network [87,90], Heap Based Optimization with Deep Quantum Neural Network [86], Quantum Relu activation function for CNN [101], Quantum Self-Supervised Network [102], 3D Quantum-inspired Self-supervised Tensor Network [103], IoTs-spiro and fuzzy-based quantum neural network system [105] |

### B. QNNs in Disease Detection and Diagnosis

A study introduced a novel quantum-inspired computational paradigm aimed at mitigating the 'dying ReLU' problem, which adversely affects the accuracy and reliability of convolutional neural networks (CNNs) using rectified linear unit (ReLU) activation functions [24,27]. By addressing critical applications such as healthcare 5.0, the proposed approach leverages quantum principles such as entanglement and superposition to derive two innovative activation functions: quantum rectified linear unit (QReLU) and modified QReLU (m-QReLU). The study demonstrated that the utilization of either QReLU or m-QReLU, in conjunction with conventional ReLU-based activation functions and their variants, led to a substantial enhancement in classification accuracy and reliability metrics for convolutional neural networks (CNNs). This improvement was observed across seven image datasets, including tasks such as Parkinson's disease detection [101]. In a research work, a quantum-inspired heuristic algorithm for blockchain-based secure healthcare prediction is presented. This novel method improves healthcare predictions' security and accuracy by fusing ideas from blockchain technology and quantum computing [67]. The use of quantum heuristic algorithms in healthcare applications was the main topic of this study on quantum models. This paper proposed a deep feedforward neural network (DFNN) optimized for healthcare forecasting using quantum krill herd optimization and the krull herd optimization (KHO) algorithm. Using blockchain technology, this system securely transmits anomalous data to servers while predicting normal or abnormal human states. It does this by guaranteeing possession, confidentiality, heterogeneity, timeliness, and incompleteness. Blockchain technology outperforms the Diffie-Hellman and RSA algorithms in terms of security by guaranteeing safe data transport to the server [94]. An excellent study has been done on how to use quantum blockchain technology to apply the Internet of Things to Healthcare 5.0. The issues of ECG data leakage in the Internet of Medical Things generation and the shortcomings of traditional blockchain technology in terms of secure storage are addressed by QADS, a quantum arrhythmia detection system that is presented in this paper. QADS improves the diagnosis of cardiovascular disease by securely storing aberrant ECG data and accurately identifying it through the use of QNN and quantum blockchain technology. To extract temporal information from ECG data, the system uses a hybrid quantum convolutional neural network (HQCNN), which has higher stability than traditional CNN models and high training and testing accuracy rates of 94.7% and 93.6%, respectively. The safety and utility of the system for managing medical data are reinforced by mathematical analysis, which verifies the robust security of the proposed quantum blockchain algorithm against various quantum attacks [93]. This analysis provides evidence that the system can effectively protect medical data from potential threats posed by quantum computing advancements. The critical requirement for an accurate and economical detection of chronic obstructive pulmonary disease (COPD), a progressive complicated illness with a significant level of mortality and morbidity, is addressed by this work. The investigation of non-invasive alternatives stems from the fact that traditional diagnostic techniques are frequently intrusive, expensive, and unreliable. With the ability to analyze breath released and detect volatile organic compounds (VOCs), the proposed IoT-Spiro System offers a promising way to diagnose COPD. Integrating this approach with a smart machine-learning prediction system improves diagnostic accuracy. It uses a hybrid genetic big bang-big Crunch method for feature extraction and a fuzzy-inspired quantum neural network (F-QNN) classifier. The IMLFF (intelligent machine learning forecasting framework) architecture is superior to current methods, as demonstrated by experimental results, indicating that the IoT-Spiro System in conjunction with IMLFF could greatly assist medical professionals in accurately diagnosing COPD [105]. In this paper, a novel model—heap-based optimization with deep quantum neural network (HBO-DQNN)—is presented that is specifically designed for healthcare data identification and classification decision-making tasks in smart healthcare systems. The HBO method is used for optimal feature selection, the DQNN model is used for data classification, and data normalization is the first step in the operation of the HBO-DQNN model. With the reported highest accuracies, the experimental findings show the effectiveness of the HBO-DQNN model and indicate its potential for improving decision-making processes in healthcare 5.0 environments [86]. The authors suggest a granular, precision-based approach to patient diagnosis, with the goal of improving illness identification and treatment accuracy and detail. This approach entails exact examinations, careful symptom reporting, in-depth professional analysis in healthcare, and a detailed knowledge of the patient's medical history. It incorporates biomolecular simulations as well as direct interventions. Using cross-validation, quantum circuit construction, probabilistic value assignment, and qubit initialization, the suggested technique makes use of quantum computing. Exhibited through a medical scenario including oxygen and heart rate levels, this methodology allows for accurate diagnosis and customized treatment based on individual biomolecular differences. On the other hand, umbrella-based strategies are less detailed and particular, frequently ignoring specific mechanisms. Early intervention may be made possible by the precision-based granular approach, which enables prompt disease identification and optimal cost efficiency [98]. A hybrid neural network is proposed by replacing the penultimate layer of the conventional neural network with a quantum layer consisting of a variational quantum circuit. In the conventional architecture, the penultimate layer is a dense layer composed of two neurons. To encode the features from the classical layer preceding the quantum layer, angle encoding is employed, which maps the features into two qubits. The $R_x$ gate provides the angle of rotation about the X-axis, which is determined by the feature's value.

$$R_x(\theta) = (\,c\,)os(\frac{\theta}{2}) - isin(\frac{\theta}{2}) - isin(\frac{\theta}{2})cos(\frac{\theta}{2}) \qquad (15)$$

Every layer rotates the system around the X-axis and then performs an entangling CNOT operation. Thus, the circuit has a total of six trainable parameters. Ultimately, the quantum state is measured and converted back into a classical output vector by the Pauli-Z operation. Fig 4 depicts the variational hybrid classical-quantum circuit's gate-level architecture [78]. To accurately predict the medication response for cancer patients (the IC50), a method that takes into account both the chemical and the cell line at the same time and predicts the outcome using a deep quantum computing circuit integrated into a machine learning architecture was developed. The resulting hybrid quantum architecture was 15% more effective than its classical equivalent in predicting the drug response. This finding offers a step toward leveraging quantum computers' capabilities to create customized medications. Combining graph convolutional, convolutional, and quantum layers results in the suggested model [51]. A quantum neural network-based multimodal fusion system for intelligent diagnosis (QNMF) is presented in a research study. It can fuse data from various modalities, handle multimodal medical data provided by IoMT devices, and enhance smart diagnosis efficiency. This system effectively extracts features from medical images using a quantum convolutional neural network (QCNN). These QCNN-based features are then utilized to train an efficient variational quantum classifier (VQC) for intelligent diagnosis by fusing them with other modality features [100]. Table III provides a detailed comparison among a set of typical quantum neural network techniques for healthcare 5.0 applications. Artificial intelligence (AI) and machine learning techniques have revolutionized the fields of life sciences and drug development, as highlighted in a specific endeavor [106]. This research utilizes generative chemistry in conjunction with near-term quantum devices to advance drug development processes. The main algorithm employed for quantum chemistry tasks is the hybrid quantum-classical Variational Quantum Eigensolver (VQE) algorithm. Generative algorithms and reinforcement learning techniques have been employed by Insilico Medicine to develop the Chemistry42 platform, which finds applications in the fields of pharmacology and chemistry. Chemistry42 is recognized as one of the most prominent software tools for de novo drug design.

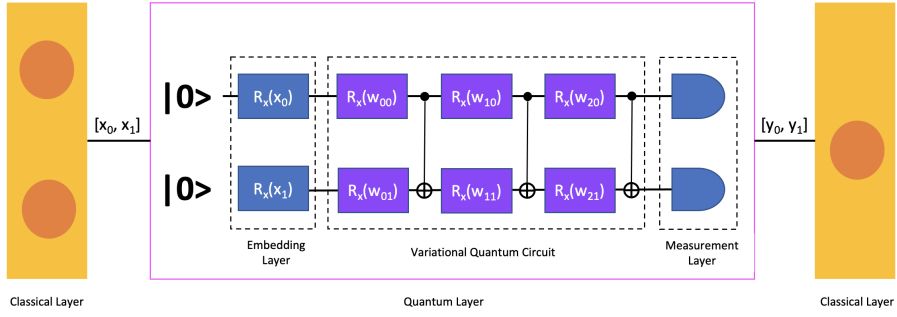

Fig. 4: Variational circuit embedded in the quantum layer

TABLE III: Comparison among typical QNN techniques for Healthcare 5.0

| Research Works | Used Databases | Technology | Experimental Setup | Accuracy (%) |
|---|---|---|---|---|
| Hybrid QNN for cancer drug prediction [51] | Genomics of Drug Sensitivity in Cancer (GDSC) database | graph convolutional, convolutional, and quantum layers | Python, IBMQiskit | 96 (outperforms by 15% from classical counterpart). |
| Classification of Brain disorder [63] | PPMI dataset, ADNI dataset | Alex-Net, quantum variational circuit and quantum layers | pennyLane simulator | Parkinson-97, Alzheimer-96 |
| Medical image classification [75] | RetinaMNIST, PneumoniaMNIST | quantum orthogonal neural networks, QNN | IBM quantum hardware, IBM Qiskit | Retina-80, Pneumonia-83 |
| Quantum Neural Network [76] | MedMNIST | Quantum encoding, quantum ansatz operation, quantum neural network | Python, IBM Qiskit | - |
| Classical-Quantum Convolutional Neural Network [78] | Alzheimer's disease dataset consisting of 6400 MRI images | variational quantum circuit, classical neural network | Python 3, PyTorch tensor, PennyLane quantum simulator, Intel Core i7 processor and 32 GB RAM and IBM Qiskit | 74.6 |
| Hybrid classical Quantum Neural Networks [81] | Alzheimer's disease dataset consisting of 6400 MRI images | modified ResNet34 architecture and a variational circuit | Python 3, TensorFlow 2, PennyLane quantum simulator, IBM Qiskit | 97.2 |
| Optimization, Deep Quantum Neural Network [86] | colon cancer and lymphoma dataset | Heap based optimization with Deep Quantum Neural Network | Python framework, IBM Q | Colon-97.05, lymphoma-95.72 |
| Smart and sustainable secure healthcare application [87] | multimedia healthcare data | explainable artificial intelligence, gradient quantum neural network, attention-based graph neural network | Python | 93 |
| Explainable MRI Radiomic QNN [89] | MRI images | quadratic unconstrained binary optimization, MRI-radiomics variational quantum neural network, quantum-annealing | PyRadiomics library version 3.0.1 on Python 3.7.9, Penny-Lane's quantum simulator | 74 |
| Quantum-LSTM contrastive learning [92] | time series mental health monitoring data | Quantum Long Short-Term Memory, Contrastive Learning framework, transfer learning | Python Jupyter notebook | 95 |
| IoMT-based smart healthcare detection system [93] | ECG heartbeat data | quantum blockchain technology, hybrid quantum convolutional neural network | - | 93.6 |
| Quantum-inspired heuristic algorithm [94] | Heart disease dataset, Thyroid Disease Dataset | Quantum-inspired heuristic algorithm combined, quantum Kril Herd Optimization, Blockchain technology | IBM-Q, Python | Heart-99.23 and Thyroid-99.97 |
| Quantum granular computing [98] | - | quantum precision-based granular approach | Falcon r5.11H processor, 1.0.24 of IBM Composer | - |
| QNMF [100] | breast cancer image data | quantum neural network, multimodal fusion | IoMT, Python, IBMQiskit | 97.07 |
| Quantum Relu activation function for Convolutional Networks [101] | MNIST dataset | Convolutional Neural Networks (CNNs), Rectified Linear Unit (ReLU) | TensorFlow and Keras, Google Colab | 99 |
| Quantum-inspired self-supervised network [102] | Dynamic Susceptibility Contrast (DSC) brain MR images | quantum-inspired network model with QM-Sig activation function | Matlab 2019 with 8 GB RAM, 3.2 GHz Processor | 99 |
| 3D Quantum-inspired Self-supervised Tensor Network [103] | 2019 Brain MR image data set and Liver Tumor Segmentation Challenge | S-connected third-order neighborhood-based topology, tensor decomposition | MATLAB 2020a, Nvidia Tesla V 100 SXM2 GPU Cluster with 32 GB of memory and 640 Tensor cores | 99.2 |
| Chronic obstructive pulmonary disease prediction [105] | Real-time dataset with 300 records, 30 features, and the target class label has two (healthy, COPD) | Genetic Algorithm (GA), Vantage Point-Tree, and Big Bang Big Crunch (BB-BC) Algorithm, Fuzzy-based Quantum Neural Network | IBM-Q, Python | 96 |

## IV. CHALLENGES AND FUTURE DIRECTIONS

Automatic medical image segmentation using deep learning frameworks often faces challenges that result in a lack of robustness and accuracy, particularly when dealing with unknown medical image classes.

- *Enhance HQNN Functionality:* The models generated in a previous study [51] will be used to construct a list of the IC50 values for specific drug/cell line pairs in the future, as well as to enhance the HQNN's functionality

- and complexity. The applicability of the HQNN to the broader healthcare environment should also be evaluated. Thus testing this architecture on other datasets linked to cancer would be helpful for the future.
- *Integrating Quantum Transfer Learning with Quantum Hardware:* The quantum transfer learning technique can eventually be used with actual quantum hardware. Moreover, computer vision multi-classification challenges can be resolved with the hybrid quantum-classical transfer learning approach [63].
- *Addressing Local Optima:* The proposed model [64] may encounter local optima due to randomization, necessitating modifications to reduce the number of features and enhance accuracy. The future work of the model will involve evaluating the categorization effectiveness of real-time data using an LSTM-based model. This approach aims to leverage the capabilities of LSTM networks to effectively analyze and classify real-time data, potentially leading to improved performance and efficiency in real-world applications.
- *Exploring Scalability and Performance:* The suggested techniques have been designed to establish a clear connection with traditional approaches, facilitating a theoretical analysis of their scalability, performance, and training time. While it is possible to combine these techniques with other forms of parametrized quantum circuits that explore larger Hilbert spaces, it remains uncertain whether and how such methods can enhance model training for conventional data. Extensive simulations and hardware experiments have been conducted, involving a substantial number of shots exceeding a billion, to showcase the capabilities and limitations of the currently available hardware [75].
- *Enhancing Performance with Larger Quantum Circuits and Massive Datasets:* To form the hybrid network, it is replaced with a single layer composed of six trainable parameters and a variational quantum circuit. Through the maturation of the software ecosystem and increased accessibility of quantum computing hardware, researchers and developers can integrate larger quantum circuits into popular classical neural networks and train them on massive datasets. In the future, researchers aim to develop a larger hybrid network that has the potential to surpass the performance of the current leading conventional networks [78].
- *Advancing MR Image Analysis:* The hybrid CNN model inspired by the ResNet34 method provides good results for the binary categorization of MR images only. In the future, researchers are interested in working on the multi-classification of MR images through a modified version of it [81].
- *Developing a Commercially Viable QML Prototype:* The goal of the authors' upcoming work is to create a fully QML-based prototype that may be commercially used in a variety of industries, including bioscience, agriculture, aviation, and life sciences. Determining an application case for quantum adaptation in real life is the primary problem [88].
- *Addressing Increased Computational Demands in Quantum Circuit:* As the required number of qubits increases, simulating the circuit classically becomes progressively more computationally demanding. However, the model architecture is designed to be scalable and can handle a larger number of features [89].
- *Enhancing Security and Precision in Healthcare Data with Quantum-Inspired Blockchain Solutions:* Ensuring data security during the application of a QNN in healthcare data is crucial. The quantum-inspired heuristic method with blockchain technology explores that option. Subsequent investigations will focus on enhancing the algorithm's efficacy concerning security and precision. The intention is to test the suggested methodology in real healthcare systems to ensure that is practical and affordable for everyday use. Moreover, future studies will investigate the suitability of the algorithm for analyzing diverse datasets related to various illnesses, such as liver-related disorders and different types of cancers [94].
- *Evaluating Quantum Activation Functions for Noisy Medical Data and Broader Clinical Applications:* Further research will concentrate on proving the enhanced performance provided by the QReLU and the m-QReLU. This will allow small clinics with limited computational resources to utilize these innovative quantum activations, thereby contributing to improved therapeutic outcomes on a broader scale. Additionally, as medical images inherently contain noise due to artifacts generated during scanning procedures and the imaging technique itself, future research will evaluate the proposed quantum activation functions in the context of noisy textual information to enhance their effectiveness in medical classification tasks [69]. This evaluation will encompass applications such as sentiment assessment and alternative model designs, including recurrent neural networks [101].
- *Challenges in Training Quantum Neural Networks for Medical Image Segmentation:* In order to effectively train a Quantum Neural Network (QNN), a labeled training set of medical images, is essential. However, the availability of medical image analysts and costly expert annotations in the datasets is limited, which hinders the full utilization of appropriate training. Moreover, the training process of quantum deep learning frameworks is time-consuming and demands expensive computational architecture and substantial memory resources. Investigating the 3D version of the QIS-Net architecture, which is relevant for volumetric MR image segmentation, is still necessary, though [102].
- *Optimizing 3D-QNet for Improved Multi-Level Segmentation Performance:* The 3D-QNet is unable to produce the best results for multi-level segmentation. It can be fixed by expanding the existing architecture, scaling up the network's interim volumetric features, and fine-tuning its hyperparameters to produce the best possible segmentation result [103].
- In the future, research will focus on utilizing the IoT-Spiro System to analyze various Volatile Organic Compound (VOC) patterns associated with different diseases.

Additionally, the effectiveness of IMLFF in diagnosing other diseases will be evaluated. The exploration of fog computing concepts, including the removal of unnecessary features at the edge of cloud storage, will also be undertaken. Furthermore, optimization strategies will be tested to enhance prediction accuracy [105].

## V. Conclusion

The current developments in quantum neural networks have enormous potential to revolutionize a number of areas of healthcare 5.0, including predictive modeling, drug discovery, medical image analysis, and operations management. Before being widely used in clinical practice, more studies are needed to address issues such as scalability, noise resilience, and interactions with the current healthcare infrastructure. This study mainly addresses the discussion of quantum computing architectures on the platform of healthcare applications. In addition, it also describes a set of popular quantum neural network systems that have multiple roles in Healthcare 5.0. This study provides the classification of several key research areas and quantum techniques in the healthcare domain along with an extensive comparison among them. Finally, this study has a range of limitations and suggests potential avenues for future research.

**Conflicts of interest**

There are no conflicts of interest.

**Data availability**

Data sharing is not applicable to this article as no new data are created or analysed in this study.

11